\documentclass[]{revtex4-2}
\usepackage{graphicx}
\usepackage{bm}
\usepackage{dcolumn}
\usepackage{amsmath}
\usepackage{amssymb}
\usepackage{epsfig}
\usepackage{color}
\usepackage[colorlinks,linkcolor=blue,citecolor=blue,urlcolor=blue]{hyperref}

\begin{document}

\title{Dynamical system analysis in descending dark energy model}

\author{M. Shahalam$^{1,2}$\footnote{E-mail address: mohdshahamu@gmail.com}}
\author{Sania Ayoub$^3$\footnote{E-mail address: saniaayoub1186@gmail.com}}
\author{Prakarshi Avlani$^{4}$\footnote{E-mail address: prakarav@gmail.com}}
\author{R. Myrzakulov$^{2}$\footnote{E-mail address: rmyrzakulov@gmail.com}}
\affiliation{$^{1}$Department of Physics, Integral University, Lucknow 226026, India}
\affiliation{$^{2}$Ratbay Myrzakulov Eurasian International Centre for Theoretical Physics \\ and Department of General \& Theoretical Physics, Eurasian National University, Nur Sultan, 010008, Kazakhstan}
\affiliation{$^{3}$School of Physical Sciences, Indian Institute of Technology, Mandi 175005, India}
\affiliation{$^{4}$Department of Physics, National Institute of Technology, Silchar 788010, India}

\begin{abstract}
In this paper, we study the dynamical system analysis for a recently proposed decaying dark energy model, namely, Q-SC-CDM. First we investigate the stationary points to find the stable attractor solution under the conditions  discussed recently in the literature. In this case, we do not find any stable attractor solution. Therefore, we avoid the parameter space of Q-SC-CDM model discussed in arXiv:2201.07704. Second, we make different choice for the model parameters and re-investigate the stationary points and their stability. Our analysis shows that a simple choice of model parameters allows to capture a stable attractor solution. Moreover, we obtain phase portrait where all trajectories move towards the stable attractor point.
\end{abstract}
\pacs{}
\maketitle

\section{Introduction}
\label{sec:intro}
Observations reveal that the universe is dominated with dark energy at the present era with an equation of state (EOS) $\omega$ close to $-1$ \cite{Planck18,Perl99,Reiss98,Sperg03,Komatsu11}. The cosmological constant is the simplest candidate of dark energy with constant EOS $-1$, and generally known as $\Lambda$CDM. However, it is plagued with fine tuning and cosmic coincidence problems. The Scalar fields are very crucial in cosmology, and has an EOS  between $+1$ and $-1$, and can easily mimic dark energy around current epoch. There is a possibility that dark energy is attributed to a type of quintessence where a scalar field evolves over time by rolling down a monotonically decreasing potential energy function $V(\phi)$. 
A wide range of various forms of the potential $V(\phi)$ has been discussed in the literature \cite{Peebles98,Wetterich88,Turner97,Cald98,Zlatev,sami06,sahni00,alam19,alam18,alam17}. The current value of $V(\phi)$ is very small that is measured in Planck mass units. This means that the energy associated with the scalar field $\phi$ is very low. There can be a potential which is shallow at present but goes to more negative values in future. A particular form of $V(\phi)$ dictates the behavior of the scalar field and its impact on the expansion of the universe. As the scalar field evolves, it can drive different phases of the universe, from the period of accelerated expansion to a slow contraction. Recently, it was shown that the quintessence models where potentials move to negative values may give rise to collapse of the universe in future that's known as Big Crunch Singularity \cite{Kal03,Garri04,Peri05,Yun04,Peri15,alam16}.

Alternatively, universe collapses or it enters into a series of cycles beginning with an expansion followed by contraction. Cyclic cosmology predicts that the universe goes through unlimited cycles, each beginning with a ``Big Bang" and ending with a ``Big Crunch". Accordingly, the universe does not start out as a singularity but rather with a ``Big Bang" and expands for a brief period of time before contraction. The principle idea behind the cyclic model is that the universe can avoid the problems related to the original singularity by undergoing infinite number of cycles, with each cycle erasing any irregularities that emerged in the preceding cycle \cite{sahni03,stein02,stein18,stein19}. For example, during early universe, the scalar field may not dominate the energy density, and the universe undergoes standard cosmological evolution, such as radiation dominance followed by matter dominance. However, when the scalar field evolves down its potential, it may eventually dominate that resulted into accelerated expansion. These transitions from matter domination to accelerated expansion, can occur smoothly and in a continuous manner dictated by the evolution of the scalar field. This allows the evolution of universe to be smooth and steady, with no abrupt or disruptive shifts. As the scalar field moves down the potential, the potential energy drops while kinetic energy increases. The kinetic energy will eventually exceed the potential energy. When this happens, the total energy density of the universe becomes zero, it means Hubble parameter, which measures the rate of expansion of the universe, also becomes zero. To this effect, the expansion of the universe stops altogether and begins to contract. The contraction phase is expected to be slow, meaning that the universe will continue to contract at a decreasing rate over a long period of time.

The dynamical system analysis is an important tool to get the asymptotic behavior of different cosmological models. Using this method, one can obtain asymptotic solutions and the stability can be checked with a simple programmed algorithm. The conventional way to get the physical cosmological solutions comes from the alliance of phase portrait and stability of critical points. The aim of this paper is to investigate the stationary points and their stability for Quintessence-Driven Slow-Contraction CDM (Q-SC-CDM) model.
The asymptotic behavior of the underlying model can be obtained through dynamical system theory. The Q-SC-CDM model was first proposed by C. Andrei $et$ $al.$ in the context of decaying dark energy and the end of cosmic expansion \cite{stein}. In our paper, we perform the phase space analysis of the model under consideration with the same model parameters as given in \cite{stein}, and investigate whether there is any stable late time de-Sitter attractor solution that leads to EOS $\omega = -1$ and energy density parameter $\Omega_{\phi}=1$. If not so, then, we shall change model parameters and re-investigate the stable attractor solutions.
The paper is organized as follows. In section \ref{sec:EOM}, we discuss the equations of motion in a homogeneous and isotropic flat Friedmann–Lemaître–Robertson–Walker (FLRW) universe, and construct the autonomous system that is useful for the dynamical system analysis. The sub-section \ref{sec:SP} is devoted to the stationary points and their stability for Q-SC-CDM model. In section \ref{sec:V0=V1},
we re-examine the fixed point analysis
 with a simple chice of model parameters. The results are concluded in section \ref{sec:conc}.

\section{Dynamics of descending dark energy model}
\label{sec:EOM}
In order to study the cosmological dynamics of a scalar field, we consider the equations of motion in a homogeneous and isotropic flat FLRW Universe as
\begin{eqnarray}
\label{eq:H}
3M_{\rm{Pl}}^2H^2 &=&\rho_m+\frac{\dot{\phi}^2}{2}+V{(\phi)}\,,\\
\label{eq:Hd}
M_{Pl}^2(2\dot H + 3H^2)&=&-\frac{\dot{\phi}^2}{2}+V(\phi)\,,\\
\label{eq:add}
{6M_{Pl}^2} \Big(\frac{\ddot a}{a}\Big) &=& -2{\dot{\phi}^2}+2V{(\phi)}-\rho_m\,,\\  
\ddot{\phi}+3H\dot{\phi}+V'(\phi) &=&0,
\label{eq:phidd}
\end{eqnarray}
where $M_{Pl}=(8 \pi G)^{-1/2}$ is the reduced Planck mass, $H=\dot{a}/a$ represents the Hubble parameter, `$a$' is the scale factor and dot denotes derivative with respect to cosmic time. The $\rho_m=\rho_{0m}a^{-3}$ designates energy density of matter with current energy density as $\rho_{0m}$.

We choose Q-SC-CDM potential proposed by C. Andrei $et$ $al.$ \cite{stein}.
\begin{equation}
V(\phi)=V_0 e^{-\phi/M}-V_1 e^{\phi/m}
\label{eq:pot}
\end{equation}
where $V_0$ and $V_1$ $(M$ and $m)$ are constants of mass dimension four (one).
We introduce the following dimensionless variables to study the dynamical system analysis.
\begin{align}
x&=\frac{\dot{\phi}}{\sqrt{6}H M_{\rm{Pl}}}\,,\quad y=\frac{\sqrt{V_0 e^{-\phi/M}}}{\sqrt{3} H M_{\rm{Pl}}}\,, \quad z=\frac{\sqrt{V_1 e^{\phi/m}}}{\sqrt{3} H M_{\rm{Pl}}}
\end{align}
In new variables the equations of motion (\ref{eq:H}) - (\ref{eq:phidd}) are written in the equivalent form of first order differential equation.
\begin{eqnarray}
x'&=x\Bigl(\frac{\ddot{\phi}}{H\dot{\phi}}-\frac{\dot H}{H^2}\Bigr),\nonumber \\
y'&=-y \Bigl(\sqrt{\frac{3}{2}}\frac{x}{M} +\frac{\dot H}{H^2}\Bigr),\nonumber\\
z'&=z \Bigl(\sqrt{\frac{3}{2}}\frac{x}{m} -\frac{\dot H}{H^2}\Bigr),
\label{eq:auton}
\end{eqnarray}
where prime ($'$) denotes derivative with respect to $\ln a$ and
\begin{align}
\frac{\dot H}{H^2}&=-\frac{3}{2} \Bigl( w_m (1-x^2-y^2+z^2) + x^2-y^2+z^2+1 \Bigr),\\
\nonumber\\
\frac{\ddot{\phi}}{H\dot{\phi}}&=-3+\sqrt{\frac{3}{2}}\frac{ y^2 M_{\rm{pl}}}{M x}+ \sqrt{\frac{3}{2}}\frac{ z^2 M_{\rm{pl}}}{m x},
\end{align}
The effective equation of state ($w_{eff}$) and equation of state ($w_{\phi}$) for field $\phi$ are given by
\begin{align}
w_{eff}&= -1 -\frac{2\dot H}{3H^2},\\
w_{\phi}&= \frac{w_{eff}-w_m \Omega_m}{1-\Omega_m},
\end{align}
where $w_m = 0$ for standard dust matter and $\Omega_m=1-x^2-y^2+z^2$. We have $w_{eff}<-1/3$ for an accelerating universe. The numerical evolution of the potential (\ref{eq:pot}), Hubble parameter and total equation of state are shown in Figs. 1 and 2 by Ref. \cite{stein}. Here, we are interested in the fixed point analysis for Q-SC-CDM model with the same conditions as chosen in Ref. \cite{stein}, see following sub-section.

\subsection{Stationary points and their stability}
\label{sec:SP}
We use autonomous system (\ref{eq:auton}) to find the stationary points by setting the left hand side of these equations to zero $(i.e.~ x'=0,~ y'=0$ and $z'=0)$. Hence, we have following stationary points.
\begin{enumerate}
\item 
\begin{eqnarray}
\label{eq:point1}
x&=& 0, \qquad y= 0, \qquad z=0,
\end{eqnarray}
The corresponding eigenvalues are
\begin{eqnarray}
{\mu}_1 &=&-3/2, \qquad
 {\mu}_2 =3/2, \qquad
  {\mu}_3 = 3/2,
\end{eqnarray}
Two eigenvalues are positive. Therefore, this is unstable point.
\item 
\begin{eqnarray}
\label{eq:point2}
x&=& \pm 1, \qquad y=0, \qquad z=0,
\end{eqnarray}
In this case, the eigenvalues are
\begin{eqnarray}
{\mu}_1 &=&3, \qquad
 {\mu}_2 =3 \mp \sqrt{\frac{3}{2}} \frac{1}{M}, \qquad
  {\mu}_3 = 3 \pm \sqrt{\frac{3}{2}} \frac{1}{m},
\end{eqnarray}
This point is also unstable as one of the eigenvalue is positive.
\item 
\begin{eqnarray}
\label{eq:point3}
x&=& 0, \qquad y= \pm \sqrt{\frac{M}{m+M}}, \qquad z=\sqrt{-\frac{m}{m+M}},
\end{eqnarray}
The eigenvalues for this point are as follows.
\begin{eqnarray}
{\mu}_1 &=&-3, \qquad
 {\mu}_2 =-\frac{3}{2} - \frac{1}{2} \sqrt{9-\frac{12}{mM}}, \qquad
  {\mu}_3 = -\frac{3}{2} + \frac{1}{2} \sqrt{9-\frac{12}{mM}},
\end{eqnarray}
The eigenvalues exhibit negative behavior for $m>0$ \& $M>0$. Hence, it is stable point. Later, we notice that the value of $z$ will be imaginary. Therefore, we are not interested in this point for $m>0$ \& $M>0$.
\item 
\begin{eqnarray}
\label{eq:point4}
x&=& -\frac{1}{\sqrt{6}m}, \qquad y= 0, \qquad z=\pm \sqrt{\frac{1-6m^2}{6m^2}},
\end{eqnarray}
We have following eigenvalues.
\begin{eqnarray}
{\mu}_1 &=&-3+ \frac{1}{2m^2}, \qquad
 {\mu}_2 =-3+ \frac{1}{m^2}, \qquad
  {\mu}_3 = \frac{m+M}{2M},
\end{eqnarray}
One of the eigenvalue is positive for $m>0$ \& $M>0$. If we choose $m<0$, one can obtain negative eigenvalues with imaginary value of  $z$, in this case, shape of potential will be changed. Hence, this point is not physically viable.
\item 
\begin{eqnarray}
\label{eq:point5}
x&=& \sqrt{3/2}M, \qquad y= \pm \sqrt{3/2}m, \qquad z=0,
\end{eqnarray}
The corresponding eigenvalues are
\begin{eqnarray}
{\mu}_1 &=&-\frac{3(1+ \sqrt{24M^2-7}}{4}, \qquad
 {\mu}_2 =-\frac{3}{4} +\frac{3(1+ \sqrt{24M^2-7}}{4}, \qquad
  {\mu}_3 = \frac{3m+3M}{2m},
\end{eqnarray}
This point is unstable for $m>0$ \& $M>0$
\item 
\begin{eqnarray}
\label{eq:point6}
x&=& \frac{1}{\sqrt{6}M}, \qquad y= \pm \sqrt{\frac{6M^2-1}{6M^2}}, \qquad z=0,
\end{eqnarray}
We have following eigenvalues
\begin{eqnarray}
{\mu}_1 &=&-3+ \frac{1}{2M^2}, \qquad
 {\mu}_2 =-3+ \frac{1}{M^2}, \qquad
  {\mu}_3 = \frac{m+M}{2mM^2},
\end{eqnarray}
Further, this point is unstable for $m>0$ \& $M>0$
\end{enumerate}
Finally, we notice that all the stationary points considered in this section (points 1-6) for Q-SC-CDM model with $m>0$ \& $M>0$ do not provide stability.


\section{ Q-SC-CDM model with  $V_0=V_1$ and $M=m$}
\label{sec:V0=V1}
In this section, we fix the model parameters as $V_0=V_1$ and $M=m$ in equation (\ref{eq:pot}), and study the asymptotic behavior of evolution equations (\ref{eq:H}) - (\ref{eq:phidd}) and phase space analysis. First we use the dynamical system approach and cast the evolution equations into an autonomous system. To do so, we introduce new dimensionless quantities defined as

\begin{align}
x&=\frac{\dot{\phi}}{\sqrt{6}H M_{\rm{pl}}}\,,\quad y=\frac{\sqrt{V}}{\sqrt{3} H M_{\rm{pl}}}\,, \quad \lambda=-M_{\rm{pl}}\frac{V'}{V}
\end{align}
Under these dimensionless quantities, an autonomous system of evolution equations takes the form
\begin{eqnarray}
x'&=&x\Bigl(\frac{\ddot{\phi}}{H\dot{\phi}}-\frac{\dot H}{H^2}\Bigr),\nonumber \\
y'&=&-y \Bigl(\sqrt{\frac{3}{2}}\lambda x+\frac{\dot H}{H^2}\Bigr),\nonumber \\
\lambda' &=&\sqrt{6}x\lambda^2(1-\Gamma),
\label{eq:auton2}
\end{eqnarray}
where prime ($'$) represents derivative with respect to $\ln a$, $\Gamma=\frac{VV_{,\phi\phi}}{V_{,\phi}^2}$ and

\begin{align}
\frac{\dot H}{H^2}&=\frac{3(y^2-x^2)}{2}-\frac{3}{2},\\
\nonumber\\
\frac{\ddot{\phi}}{H\dot{\phi}}&=-3+\sqrt{\frac{3}{2}}\frac{ \lambda y^2}{x},
\end{align}
The equation of state for the field $\phi$ is given as,
\begin{align}
w_{eff}&= -1 -\frac{2\dot H}{3H^2},\\
w_{\phi}&= \frac{w_{eff}-w_m \Omega_m}{1-\Omega_m},
\end{align}
where $w_m = 0$ for standard dust matter and $\Omega_m=1-x^2-y^2$. 
Since, 
\begin{equation}
\Omega_m+\Omega_{\phi}=1
\end{equation}
Therefore, it provides the constraint as
\begin{equation}
0 \leq \Omega_{\phi}= x^2+y^2  \leq 1
\end{equation}
Under the chosen parameters
\begin{equation}
\Gamma=\frac{VV_{,\phi\phi}}{V_{,\phi}^2}= \frac{1}{m^2\lambda^2}
\label{eq:Gamma}
\end{equation}

\begin{figure}
\begin{center}
\begin{tabular}{c }
{\includegraphics[width=3in,height=3in,angle=0]{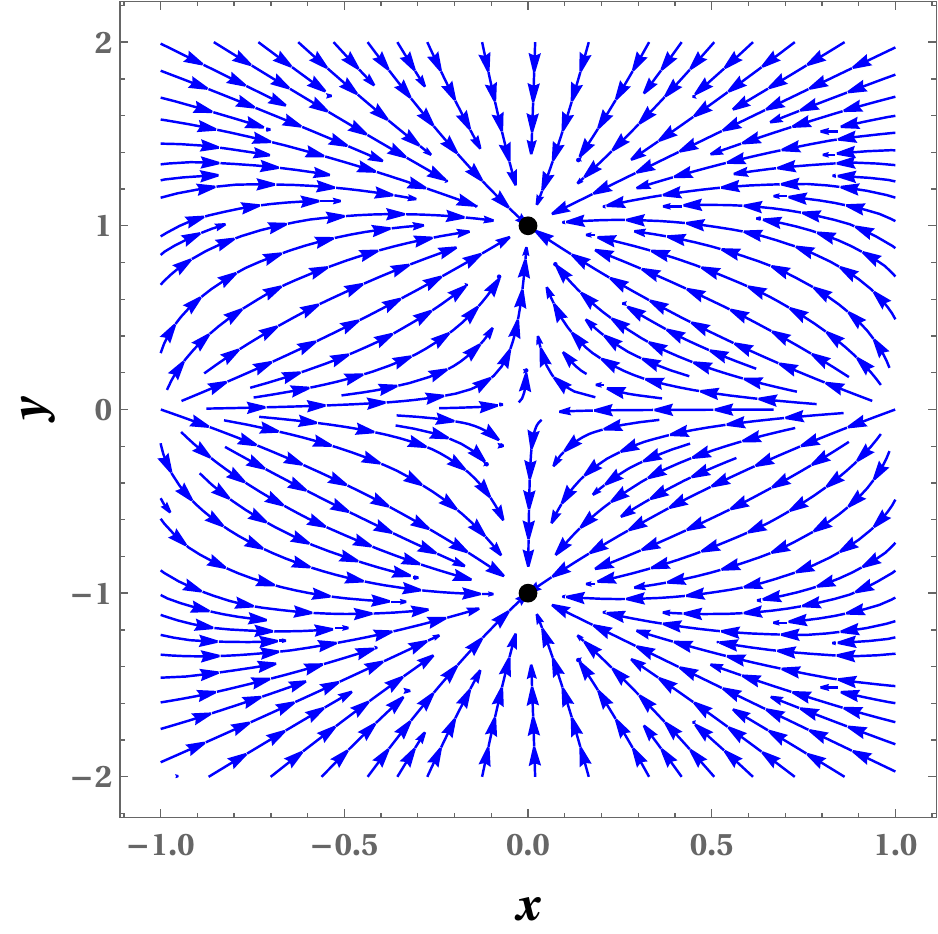}}
\end{tabular}
\caption{The figure exhibits phase space trajectories for point 2 with $m=1$ in $x-y$ plane. The stable point behaves as an attractive node for which $w_{eff}= w_{\phi}= -1, \Omega_{\phi}=1 ~\text{and}~ \Omega_m=0$. The black dot shows stable attractor point where all the trajectories meet.}
\label{fig:port}
\end{center}
\end{figure}

We numerically evolve autonomous system (\ref{eq:auton2}) with (\ref{eq:Gamma}), the stationary points and eigenvalues are given as 
\begin{enumerate}
\item 
\begin{eqnarray}
\label{eq:point1a}
x&=& 0, \qquad y= 0,  \qquad \lambda= \pm \frac{1}{m}
\end{eqnarray}
The corresponding eigenvalues are
\begin{eqnarray}
{\mu}_1 &=&-3/2, \qquad
 {\mu}_2 =3/2, \qquad
  {\mu}_3 = 0,
\end{eqnarray}
This is unstable point.
\item 
\begin{eqnarray}
\label{eq:point2a}
x&=& 0, \qquad y= \pm 1, \qquad \lambda=0,
\end{eqnarray}
We have following eigenvalues
\begin{eqnarray}
{\mu}_1 &=&-3, \qquad
 {\mu}_2 = -\frac{3}{2} + \frac{\sqrt{9m^2-12}}{2m}, \qquad
  {\mu}_3 = -\frac{3}{2} - \frac{\sqrt{9m^2-12}}{2m}
\end{eqnarray}
The eigenvalues are negative for all values of $m$ provided that $m \neq 0$. Therefore, this critical point is stable point. In this case, the equation of state $w_{eff}= w_{\phi}= -1$ and energy density parameters $\Omega_{\phi}=1 ~\text{and}~ \Omega_m=0$.
\item 
\begin{eqnarray}
\label{eq:point3a}
x&=& \pm \sqrt{\frac{3}{2}}m, \qquad y=  \mp \sqrt{\frac{3}{2}}m, \qquad \lambda= \pm \frac{1}{m}
\end{eqnarray}
The eigenvalues are given as
\begin{eqnarray}
{\mu}_1 &=&6, \qquad
 {\mu}_2 = - \frac{3 (\sqrt{24m^2-7}+1)}{4}, \qquad
  {\mu}_3 = \frac{3 (\sqrt{24m^2-7}+1)}{4}
\end{eqnarray}
This point is unstable as two eigenvalues are positive.
\item 
\begin{eqnarray}
\label{eq:point4a}
x&=& \pm \frac{1}{\sqrt{6}m}, \qquad y= \pm \sqrt{1-\frac{1}{6m^2}}, \qquad \lambda= \pm \frac{1}{m}
\end{eqnarray}
In this case, the eigenvalues are 
\begin{eqnarray}
{\mu}_1 &=& 2, \nonumber \\
 {\mu}_2 &=& \frac{2-13m^2+2m^4-\sqrt{9-62m^2+75m^4-22m^6+m^8}}{4m^2}, \nonumber \\
  {\mu}_3 &=& \frac{2-13m^2+2m^4+\sqrt{9-62m^2+75m^4-22m^6+m^8}}{4m^2}
\end{eqnarray}
This point is also unstable as one of the eigenvalue is positive.

In this section, only critical point 2 shows stability with eigenvalues ${\mu}_1 = -3, {\mu}_2 = (-3+ i \sqrt{3})/2$ and ${\mu}_3 = (-3- i \sqrt{3})/2$ for $m=1$, and behaves as an attractive node. One can also find stability for other values of $m$ except $m=0$. The phase portrait for stable point 2 is depicted in Fig. \ref{fig:port} where all the trajectories of phase portrait go towards the stable attractor point. The stable point represents the potential energy dominated behavior with energy density parameters $\Omega_{\phi}=1, \Omega_m=0$ and EOS $w_{eff}= w_{\phi}= -1$.
\end{enumerate}

\section{Conclusion}
\label{sec:conc}
In our work, we used Q-SC-CDM model to investigate the dynamical system analysis. First, we constructed an autonomous system for said model to perform the phase space analysis. Second, we obtained the critical points and discussed their stabilities by looking the eigenvalues, under the same conditions $(i.e.~ m>0 ~\&~ M>0)$ of Ref. \cite{stein}. It was noticed that none of the critical points under the decaying dark energy conditions exhibit stable attractor solution. Let us discuss critical points 3 \& 4 of sub-section \ref{sec:SP} briefly. The point 3 showed stability under $m>0 ~\&~ M>0$, later we see that the dimensionless variable $z$ will be imaginary. Similarly, point 4 presented negative eigenvalues for $m<0 ~\&~ M>0$ (for this condition, shape of the potential will be changed), again $z$ will be imaginary. Therefore, both the points are not viable physically. Next, to search the stable attractor solution, we made a simple change in the model parameters of Q-SC-CDM model as $V_0=V_1$ \& $M=m$, and re-investigated the phase space analysis in section \ref{sec:V0=V1}. In this case, the critical point 2 showed stability with eigenvalues  ${\mu}_1 = -3, {\mu}_2 = (-3+ i \sqrt{3})/2$ and ${\mu}_3 = (-3- i \sqrt{3})/2$ for $m=1$, and behaves as an attractive node with $w_{eff}= w_{\phi}= -1$ and  $\Omega_{\phi}=1$. The phase portrait is presented in Fig. \ref{fig:port} where all trajectories move towards stable attractor point.


\section*{Acknowledgments}
We are indebted to M. Sami for assigning the project, useful discussions and comments. Author MS thanks the Inter-University Centre for Astronomy and Astrophysics (IUCAA), Pune for the hospitality and facilities under the visiting associateship program where the part of the work was done. MS would also like to acknowledge Swagat Saurav Mishra for the valuable discussion. This work was partially supported by the Ministry of Science and Higher Education of the Republic of Kazakhstan, Grant No. AP14870191.

\end{document}